\documentclass[twocolumn,showpacs,preprintnumbers,amsmath,amssymb,prb]{revtex4}

\usepackage{graphicx}
\usepackage{dcolumn}
\usepackage{bm}

\begin{document}


\title{Unusual electron-doping effects in Sr$_{2-x}$La$_x$FeMoO$_6$ observed by photoemission spectroscopy}

\author{T. Saitoh}
 \altaffiliation[Present Address: ]{Department of Applied Physics, Tokyo University of Science, Shinjuku-ku, Tokyo 162-8601, Japan}
 \email{t-saitoh@rs.kagu.tus.ac.jp}
\author{M. Nakatake}
 \altaffiliation[Present Address: ]{Hiroshima Synchrotron Radiation Center, Hiroshima University, Higashi-Hiroshima, Hiroshima 739-8521, Japan.}
\affiliation{Photon Factory, Institute of Materials Structure Science, KEK, Tsukuba, Ibaraki 305-0801, Japan}

\author{H. Nakajima}
 \altaffiliation[Present Address: ]{National Synchrotron Research Center, PO Box 93, Nakhon-Ratchasima 30000, Thailand.}
\author{O. Morimoto}
\affiliation{Department of Materials Structure Science, Graduate University for Advanced Studies, Tsukuba, Ibaraki 305-0801, Japan}

\author{A. Kakizaki}
\affiliation{Institute for Solid State Physics, University of Tokyo, Kashiwa, Chiba 277-8581, Japan}

\author{Sh. Xu and Y. Moritomo}
\affiliation{Department of Applied Physics, Nagoya University, Nagoya 464-8603, Japan}

\author{N. Hamada}
\affiliation{Department of Physics, Tokyo University of Science, Chiba 278-8510, Japan}

\author{Y. Aiura}
\affiliation{National Institute of Advanced Industrial Science and Technology, Tsukuba, Ibaraki 305-8568, Japan}

\date{\today}

\begin{abstract}
We have investigated the electronic structure of electron-doped Sr$_{2-x}$La$_x$FeMoO$_6$ ($x$=0.0 and 0.2) by photoemission spectroscopy and band-structure calculations within the local-density approximation+$U$ (LDA+$U$) scheme. A characteristic double-peak feature near the Fermi level ($E_{\rm F}$) has been observed in the valence-band photoemission spectra of both $x$=0.0 and 0.2 samples. 
	A photon-energy dependence of the spectra in the Mo 4$d$ Cooper minimum region compared with the band-structure calculations has shown that the first peak crossing $E_{\rm F}$ consists of the (Fe+Mo) $t_{2g\downarrow}$ states (feature A) and the second peak well below $E_{\rm F}$ is dominated by the Fe $e_{g\uparrow}$ states (feature B). 
	Upon La substitution, the feature A moves away from $E_{\rm F}$ by $\sim$50 meV which is smaller than the prediction of our band theory, 112 meV. In addition, an intensity enhancement of $both$ A and B has been observed, although B is not crossing $E_{\rm F}$. Those two facts are apparently incompatible with the simple rigid-band shift due to electron doping. We point out that such phenomena can be understood in terms of the strong Hund's rule energy stabilization in the 3$d^5$ configuration at the Fe sites in this compound.
	From an observed band-narrowing, we have also deduced a mass enhancement of $\sim$2.5 with respect to the band theory, in good agreement with a specific heat measurement.
\end{abstract}

\maketitle

\section{\label{sec:level1}Introduction}

Industrial demands on seeking new materials with exotic magneto-transport properties have been expanding the basic research field of transition-metal oxides with unusual magnetic and transport properties.
	Recent re-investigations on the family of double perovskite-type oxides $A_2BB'$O$_6$ are one of such examples. The revived interest on the double perovskites has its origin in the large tunneling magneto-resistance discovered in Sr$_2$FeMoO$_6$ and Sr$_2$FeReO$_6$,\cite{KobayashiNature,KobayashiRe} although it has already  been known since 1960's that Sr$_2$FeMoO$_6$ is a ferrimagnetic (or ferromagnetic) metal with a quite high ferrimagnetic transition temperature ($T_{\rm C}$) of 420 K.\cite{Galasso} 

	Several band-structure calculations and optical or electron-spectroscopic studies have confirmed that those iron-based compounds generally have the half-metallic density of states (DOS) at the Fermi level ($E_{\rm F}$).\cite{KobayashiNature,MoritomoPRB,Tomioka,Fang,Wu,Kang,SaitohDP}
	Ferrimagnetism accompanied by metallic conductivity and the half-metallic density of states (DOS) naturally reminds us of the colossal magnetoresistive manganates and the double exchange (DE) mechanism. Indeed, several authors have pointed out that DE can explain the electronic properties of Sr$_2$FeMoO$_6$,\cite{Kang,SaitohDP,Martinez,MoritomoPRBR} while others have proposed a new mechanism of ferrimagnetic metal.\cite{SarmaPRL,KT}
	Nevertheless, it is common in any models that the carrier density or DOS at $E_{\rm F}$ has much importance since the ferromagnetic interaction between Fe local spins is mediated by charge carriers (in DE models) or the Fe-O-Mo hybridized states (in hybridization models).

	In this sense, a study of carrier-doping effects on the electronic structure of Sr$_2$FeMoO$_6$ is necessary to seek the origin of the ferrimagnetism of this compound. Navarro {\it et al.} have recently investigated this issue using polycrystalline samples of Sr$_{2-x}$La$_x$FeMoO$_6$.\cite{NavarroPES} Sr$_{2-x}$La$_x$FeMoO$_6$ can be regarded as an electron-doped system of Sr$_2$FeMoO$_6$, where $x$ corresponds to the number of doped electron per one Fe/Mo site. They have found that the $E_{\rm F}$ spectral weight linearly increases with $x$, in accordance with a linear increase of $T_{\rm C}$.
	Although their result and argument seem to be clear and reasonable, there still remains an experimental and a theoretical concern: the former is about polycrystalline nature of their samples as well as the scratching surface treatment. In our previous paper,\cite{SaitohDP} we have intensively discussed this issue and shown that spectra of Sr$_2$FeMoO$_6$ from a scraped and a fractured surface were quite different. In particular, the near-$E_{\rm F}$ intensity was found to be considerably suppressed in scraping measurements. Since the near-$E_{\rm F}$ intensity is directly relevant to which model is plausible, a study using single crystals should be needed to address the above issue.
	In connection with theoretical studies, on the other hand, electron-doping effects should be examined first by band theory before we consider the DE or other new models. For example, if the calculated DOS is increasing monotonically with electron energy, the $E_{\rm F}$ DOS should linearly increase with $x$. 

	In this paper, we study the electron-doping effects on the electronic structure of single-crystalline Sr$_2$FeMoO$_6$ to give insight into the mechanism of ferrimagnetism and the half-metallic DOS by photoemission spectroscopy combined with LDA+$U$ band-structure calculations. To avoid possible complications rising from a structural phase transition and anti-site effects in heavily doped region\cite{Navarro2001} and to probe only the electronic effects due to La substitution, we concentrate on a lightly-doped region.

\section{Experiment and Calculation}

	High quality single crystals of Sr$_{2-x}$La$_x$FeMoO$_6$ ($x$=0.0 and 0.2) were grown by floating-zone method.\cite{MoritomoPRB} The site disorder was about 10$-$15\% which lowers the $T_{\rm{C}}$ from the ideal values,\cite{MoritomoJJAP2001}  but will not seriously affect the microscopic electronic structure.\cite{SaitohDP}  The experiments have been performed at the beamline BL-11D of the Photon Factory using a Scienta SES-200 electron analyzer. The total energy resolution was about 50$-$90 meV FWHM using 65$-$200 eV photon energies. The vacuum was always better than 1.5$\times$10$^{-10}$ Torr and the temperature was about 20 K. 
	To obtain the best quality of surface, we have fractured samples {\it in situ} at 20 K. The prepared surface was blackly shining like a cleaved surface, but was rough enough to get angle-integrated spectra although angle-resolved effects appeared in low photon energies to some extent. For comparison, we have also scraped samples with a diamond file.\cite{SaitohDP} The spectral intensity was normalized by the total area of the full valence-band spectra and the near-$E_{\rm F}$ spectra were scaled to them.

  Band-structure calculations for non-doped Sr$_2$FeMoO$_6$ have been performed with the full-potential linearized augmented plane-wave (FLAPW) method within the local-density approximation (LDA)+$U$ scheme. For effective Coulomb repulsions $U_{\rm eff}=U-J$, we have adopted rather small values (2.0 eV for Fe and 1.0 eV for Mo, respectively). More detailed information is given in Ref.~\onlinecite{SaitohDP}.

\section{Results and discussion}

\begin{figure}[b]
	\begin{center}
	\includegraphics[width=75mm,keepaspectratio]{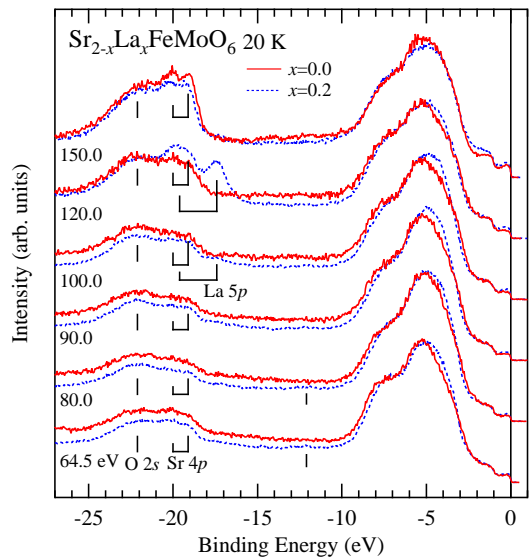}
	\end{center}
		\caption{(Color online.) Photoemission spectra of the valence-band with shallow core levels of Sr$_{2-x}$La$_x$FeMoO$_6$ at 20K.}
	\label{figWide}
\end{figure}

	Figure~\ref{figWide} shows photoemission spectra of the valence-band of Sr$_{2-x}$La$_x$FeMoO$_6$ with shallow core levels at 20K. A doublet structure at $-19.1$ and $-20.0$ eV is the Sr 4$p$ core level, which is on the long tail of the O 2$s$ core level at $-22.1$ eV. Upon La substitution, another doublet structure due to the La 5$p$ core level appears at -17.4 and -19.6 eV. However, it is very weak due to a small photoionization cross section.\cite{Yeh} The strong enhancement of the La 5$p$ intensity at $h\nu$=120 eV is attributed to the La 4$d-$4$f$ giant resonance.\cite{Molodtsov} A very small structure at 12.1 eV observed in lower photon-energy spectra is most likely due to surface aging effects.

\begin{figure}[]
	\begin{center}
	\includegraphics[width=75mm,keepaspectratio]{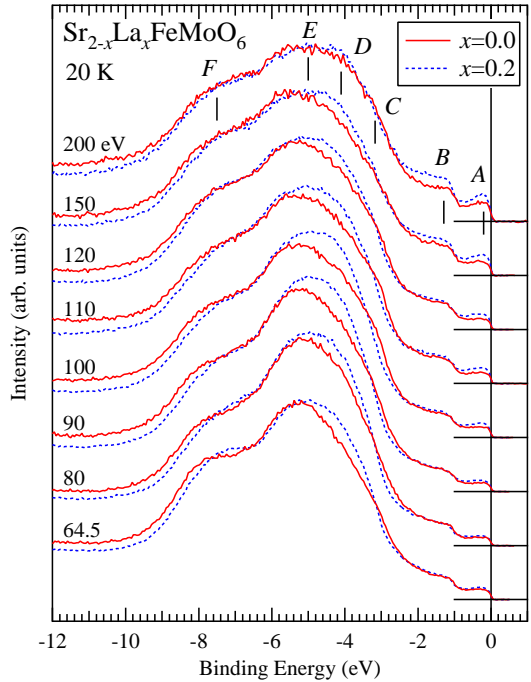}
	\end{center}
		\caption{(Color online.) Valence-band photoemission spectra of Sr$_{2-x}$La$_x$FeMoO$_6$ at 20K.}
	\label{figVB}
\end{figure}

	Figure~\ref{figVB} shows full valence-band spectra of Sr$_{2-x}$La$_x$FeMoO$_6$ at 20 K. One can observe six structures denoted as $A$ to $F$. A comparison with the band-structure calculations clarifies that the double-peak structures $A$ and $B$ near $E_{\rm F}$ correspond to the (Fe+Mo) $t_{2g\downarrow}$ and Fe $e_{g\uparrow}$ bands, respectively.\cite{KobayashiNature,Fang,Wu,Kang,SaitohDP,MoritomoPRBR,SarmaPRL} Also, $C$ and $D$ mostly originate from the Fe $t_{2g\uparrow}$ bands with a contribution from the O 2$p$ intensity. $E$ is predominantly due to the O 2$p$ non-bonding states. The Fe $t_{2g\uparrow}$ and $e_{g\uparrow}$ bonding states contribute to $F$ to some extent.
	The features $D$ and $E$ are somewhat enhanced in the low photon-energy spectra upon La substitution. This can be primarily interpreted as angle-resolved effects because the enhancement becomes small with increasing photon energy and almost vanishes for all $C$$-$$F$ at the highest 200 eV spectrum.

\begin{figure*}[]
	\begin{center}
	\includegraphics[width=135mm,keepaspectratio]{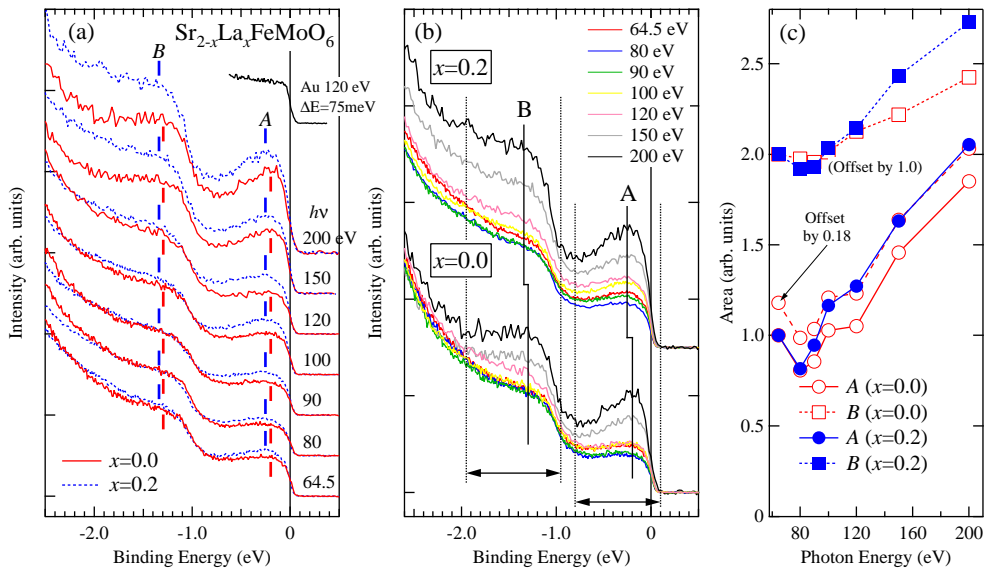}
	\end{center}
		\caption{(Color.) (a) Near-$E_{\rm F}$ photoemission spectra of Sr$_{2-x}$La$_x$FeMoO$_6$ at $T$=20K. Red ($x$=0.0) and blue ($x$=0.2) bars indicate the locations of the feature $A$ and $B$. A gold spectrum at 120 eV is also shown.(b) Photon-energy dependence of the $x$=0.0 and the $x$=0.2 near-$E_{\rm F}$ spectra. The data are same as (a). (c) Normalized spectral weight of $A$ and $B$ calculated using windows of $-0.8$ to $0.1$ eV ($A$) and $-1.95$ to $-0.95$ eV ($B$). The windows are indicated in Panel (b). The spectral weight at 64.5 eV is set to unity. Note that the weight for $B$ includes an offset of 1.0.}
	\label{figEF}
\end{figure*}

\begin{figure}[b]
	\begin{center}
	\includegraphics[width=75mm,keepaspectratio]{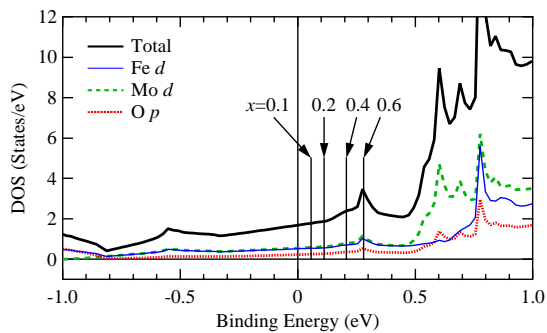}
	\end{center}
		\caption{(Color online.) Total and partial DOS of Sr$_2$FeMoO$_6$ in the near-$E_{\rm F}$ region calculated by the LDA+$U$ method. Vertical lines denote expected locations of $E_{\rm F}$ deduced from the calculation, assuming the rigid band shift due to electron doping.}
	\label{figBandShift}
\end{figure}

	By contrast, substantial changes are observed in the near-$E_{\rm F}$ region of all the spectra as shown in Fig.~\ref{figEF}. Panel (a) of Fig.~\ref{figEF} shows near-$E_{\rm F}$ spectra of Sr$_{2-x}$La$_x$FeMoO$_6$. Upon La substitution, the features $A$ and $B$ are shifted from $-0.20$ to $-0.25$ eV and from $-1.30$ to $-1.34$ eV, respectively. Although the direction of the shift is in accordance with electron doping, the amount of the shift ($\sim$40$-$50 meV)\cite{note5}  is too small; Figure~\ref{figBandShift} illustrates the expected location of $E_{\rm F}$ deduced from our LDA+$U$ band-structure calculation assuming the rigid band shift. It predicts that $x$=0.2 (0.2 electron doping per one Fe/Mo site) should correspond to a $E_{\rm F}$ shift of 112 meV while the 50 meV shift of $E_{\rm F}$ corresponds to $x$=0.086. 

	Panel (b) of Fig.~\ref{figEF} shows a photon-energy dependence of the near-$E_{\rm F}$ spectra. The intensity of the features $A$ and $B$ tends to increase with $h\nu$, indicating that considerable Fe 3$d$ weight compared to the O 2$p$ one exists in those features.\cite{Yeh} However, one can notice that the intensity at $A$ does not increase monotonically but has a minimum at $\sim$80 eV while such a clear minimum is not observed in $B$. This is because the Cooper minimum of Mo 4$d$ states strongly suppresses the Mo 4$d$ weight around $\sim$80$-$90 eV and only the feature $A$ has a substantial contribution from Mo 4$d$ states.\cite{SaitohDP,Yeh}

	 These behaviors of the spectral weight of $A$ and $B$ vs. $h\nu$ are summarized in Panel (c). Panel (c) shows a normalized spectral weight of $A$ and $B$. Here we set the 64.5 eV to unity as a reference.
	A clear minimum around 80 eV for the feature $A$ is attributed to the Cooper minimum of Mo 4$d$ states. The Cooper minimum is obviously enhanced in the $x$=0.2 curve while the two curves are virtually parallel to each other above the minimum. This observation indicates that the Mo 4$d$ contribution to the feature $A$ is larger for $x$=0.2 than $x$=0.0, but no significant change in the Fe 3$d$ and O 2$p$ contributions. Namely, the doped electrons are mainly introduced into the Mo 4$d$ $t_{2g}$ states, as inferred by Moritomo {\it et al.}\cite{MoritomoPRB} More recently, Frontera {\it et al.} have observed by neutron diffraction that the Mo$-$O distance increases with La doping while the Fe$-$O one does not change.\cite{Frontera} In terms of the ionic-radius argument, this implies that the doped electrons will mainly be located at the Mo sites,\cite{Frontera} in agreement with the above argument.
	On the other hand, the spectral weight at the feature $B$ of $x$=0.2 is also considerably enhanced in the high photon energies despite the fact that the two curves are virtually identical below 120 eV. This is indicating that the feature $B$ (Fe $e_{g\uparrow}$ bands) also obtains electrons. However, it cannot be a simple consequence of electron doping because the feature $B$ is not crossing over $E_{\rm F}$ owing to the half-metallic DOS.

	Such an unusual behavior, the enhancement of {\it both} $A$ and $B$ due to electron doping, has also been reported in the recent photoemission study on polycrystalline samples by Navarro {\it et al.}\cite{NavarroPES} The fact that two independent experiments using different samples with different surface treatments have given the same result apparently demonstrates that this is an intrinsic change of the electronic structure due to electron doping. 
	This is of course not of the rigid-band type, but also incompatible with the behavior of typical electron- or hole-doped 3$d$ transition-metal oxides such as La$_{1-x}$Sr$_x$TiO$_3$ or La$_{1-x}$Sr$_x$MnO$_3$. In these compounds, in-gap states induced by carrier doping always appear between the top of the valence band and the bottom of the conduction band and are crossing $E_{\rm F}$.\cite{FujimoriLSTO,MorikawaYCTO,SaitohMn,JHPMn}

	We believe that the above strange behavior can be understood in terms of the strong Hund's rule coupling in the Fe $d^5$ configuration as follows: the electron configuration in Sr$_2$FeMoO$_6$ is not completely $[3d^5+4d^1]$-like\cite{note6} but still has some weight of the $[3d^5\underline{L}(e_g)+4d^2]$ configuration in which $3d^5\underline{L}(e_g)$ is the dominant electron configuration in the ``original" SrFeO$_3$-like environment for Fe ions.\cite{SaitohDP} Here, $\underline{L}$ denotes an O 2$p$ ligand hole. Upon electron doping, the electron configuration will be changing from either $[3d^5+4d^1]$ or $[3d^5\underline{L}(e_g)+4d^2]$ configuration to be more like a $[3d^5+4d^2]$ configuration because the strong Hund's rule energy stabilization of the $d^5$ configuration prevents the Fe sites from having more than five electrons.\cite{SaitohDP} As a consequence, the doped electrons will occupy either Mo 4$d$ states or ligand-hole states, resulting in the enhancement of both features $A$ (Mo 4$d$ states) and $B$ (Fe 3$d$ $e_g$ states).\cite{note1}
	The enhancement in the feature $B$ due to electron doping, thus, reflects the strong Hund's rule energy stabilization in this compound.

\begin{figure}[b]
	\begin{center}
	\includegraphics[width=75mm,keepaspectratio]{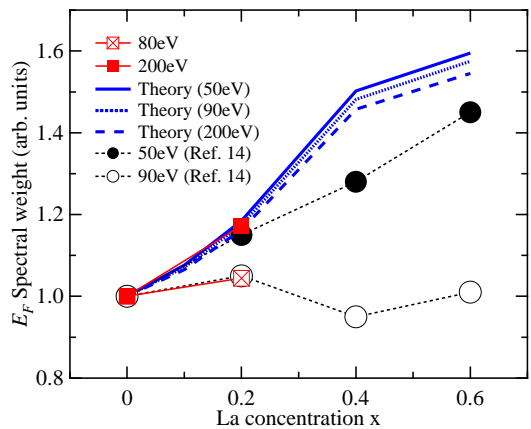}
	\end{center}
		\caption{(Color online.) Experimental and theoretical $E_{\rm F}$ spectral weight plotted as functions of La concentration $x$. The $E_{\rm F}$ spectral weight was evaluated from integration within a $\pm$0.1 eV window at $E_{\rm F}$ and normalized with respect to the $x$=0.0 one. The rigid-band shift is assumed in the theoretical curves and the photoionization cross sections of Fe 3$d$, Mo 4$d$ and O 2$p$ states are taken into accounted. Results by Navarro {\it et al.} (Ref.\protect\onlinecite{NavarroPES}) are also shown for comparison.}
	\label{figWeight}
\end{figure}

	Figure~\ref{figWeight} shows a comparison between experimental and theoretical $E_{\rm F}$ spectral weight plotted as functions of La concentration $x$. The photoionization cross sections are taken into accounted for theoretical curves. The theoretical $E_{\rm F}$ weight almost linearly increases with $x$. 
	Here, the 50 eV and 200 eV curves predominantly represent both Fe 3$d$ and Mo 4$d$ weight and the 90 eV should represent only the Fe 3$d$ weight.\cite{note3}
	It is noted that all the three theoretical curves have no significant difference although the 90 eV curve would have smaller weight due to the Mo 4$d$ Cooper minimum. This comes from the small Mo 4$d$ DOS due to the high valence of Mo ions.
	Nevertheless, the experimental $E_{\rm F}$ weight for 80 eV and 200 eV displays a considerable difference.
	Here it is worthy to note that our 200 eV and 80 eV curves are almost identical to the 50 eV and 90 eV ones by Navarro {\it et al.}, respectively.\cite{note4} 
	On the other hand, they have reported a linear relationship between the $E_{\rm F}$ spectral weight and $T_{\rm C}$ (see Fig.~\ref{figWeight}).\cite{NavarroPES}  One may consider that it can be an evidence of the DE mechanism in this compound although they carefully mentioned that there were several possibilities.
	In our measurements, $T_{\rm C}$ is not enhanced upon electron doping\cite{MoritomoPRB} due to the site disorder.\cite{MoritomoJJAP2001} It is noted, however, that the observed $E_{\rm F}$ spectral weight is almost identical to their results. Therefore we infer that the enhancement of the $E_{\rm F}$ weight in both experiments may not be a direct consequence of the DE mechanism although we also believe that the DE mechanism can basically describe the electronic structure of this compound.\cite{SaitohDP}

	The discrepancy between theory and experiment in the Cooper minimum region possibly indicates that the Mo 4$d$ states have larger near-$E_{\rm F}$ weight than expected from the band theory. It does not necessarily mean that more Mo 4$d$ electrons nominally exist but the O 2$p$ and Fe 3$d$ $t_{2g\downarrow}$ states, which strongly hybridize with Mo 4$d$ $t_{2g\downarrow}$ ones, may be able to contribute to the Mo 4$d$ spectral weight to some extent. This argument can also explain a small (a factor of 1.5) suppression of the intensity at the feature A in an experimental spectrum compared with a band theory simulation shown in Fig.~5 of Ref.~\onlinecite{SaitohDP}. However, it would be inconsistent with M\"ossbauer measurements which concluded Fe$^{2.5+}$.\cite{Linden,Nakamura} Hence this discrepancy is an open question at this stage.

	On the other hand, both our 200 eV curve and the 50 eV curve by Navarro {\it et al.} coincides with the theoretical ones up to $x$=0.2. Because we set all the spectral weights for $x$=0.0 to unity as a reference, this implies $W_{\rm exp}(0.2)/W_{\rm exp}(0.0) \approx W_{\rm B}(0.2)/W_{\rm B}(0.0)$, where $W_{\rm exp}$ and $W_{\rm B}$ denote the experimental and theoretical spectral weight at $E_{\rm F}$, respectively.
	However, this is not a consequence of the rigid-band shift, as we have presented above. Instead, this situation can be realized if a band-narrowing occurs uniformly in the (Fe+Mo) $t_{2g\downarrow}$ band. We believe that such a band-narrowing is realized since the system is a good metal. In La$_{1-x}$Sr$_x$TiO$_{3+y/2}$ case, for example, this type of narrowing appears in the Fermi liquid phase whereas a narrowing occurs only in the vicinity of $E_{\rm F}$ when the system is close to the metal-insulator transition.\cite{YoshidaLTO} 

	Based upon the uniform band-narrowing assumption, the ratio of the theoretical chemical potential shift to that of the experiment should rerpresent the mass enhancement from the band mass. In Table~\ref{Table1}, two estimations of mass enhancement $\gamma_{\rm exp}/\gamma_{\rm B}$ and $W_{\rm exp}/W_{\rm B}$ are listed. 
	A theoretical electronic specific heat $\gamma_{\rm B}$ is deduced from the band-theory DOS at $E_{\rm F}$ [$N_{\rm B}(E_{\rm F})$] using the formula 
	$\gamma_{\rm B} = \pi^2k_B^2 N_{\rm B}(E_{\rm F}) / 3$.
$W_{\rm exp}/W_{\rm B}$ describes a mass enhancement estimated from the band-narrowing. For $x$=0.0, this number is deduced from the location of the feature $A$ (theory: $-0.50$ eV, experiment: $-0.20$ eV)\cite{SaitohDP} and for $x$=0.2, we make use of the ratio of the chemical potential shift (theory: 112 meV, experiment: 40$-$50 meV) based on the above argument. They give $W_{\rm exp}/W_{\rm B}$ of 2.5 ($x$=0.0) and 2.2$-$2.8 ($x$=0.2), in good agreement with the estimation from $\gamma_{\rm exp}$.
	Therefore, electron-doping effects on the (Fe+Mo) $t_{2g\downarrow}$ band can be understood in terms of a conventional electron doping into a renormalized (by a factor of two) band like the La$_{1-x}$Sr$_x$TiO$_3$ case.\cite{YoshidaLTO,TokuraPRL} Currently, we have no idea to determine how many electrons will be introduced into the Mo(+Fe) $t_{2g\downarrow}$ band and the Fe $e_{g\uparrow}$ band, respectively, in our scenario of electron doping: $\alpha|$$3d^5 4d^1$$>$ + $\beta|$$3d^5\underline{L}(e_g)4d^2$$>$ $\rightarrow$ $|$$3d^54d^2$$>$ ($|\alpha| \gg |\beta|$). However, it is safe to say that the doping effects should appear in the feature $A$ more than in the feature $B$ because $|\alpha| \gg |\beta|$. In this sense, the observed enhancement in the feature $A$ is rather smaller than expected. This can be understood again in connection with the small photoionization cross section of the Mo 4$d$ states, if we assume that the electrons doped into the (Fe+Mo) $t_{2g\downarrow}$ band will mostly occupy the Mo 4$d$ states.

\begin{table}
\caption{\label{Table1}LDA+$U$ band DOS at $E_{\rm F}$ $N_{\rm B}(E_{\rm F})$ (in 10$^{24}$ eV$^{-1}$mol$^{-1}$), electronic specific heat $\gamma_{\rm B}$ deduced from $N_{\rm B}(E_{\rm F})$ [in mJ/(K$^2$mol)], a mass enhancement estimated from the electronic specific heat $\gamma_{\rm exp}/\gamma_{\rm B}$, and a mass enhancement estimated from the  band-narrowing $W_{\rm exp}/W_{\rm B}$. }
\begin{ruledtabular}
\begin{tabular}{cccccc}
$x$ 											&0.0		&0.1		&0.2		&0.4		&0.6\\
\tableline
$N_{\rm B}(E_{\rm F})$\footnote{Rigid-band shift is assumed for $x>0$.}	
												&1.0		&1.1		&1.1		&1.4		&1.9\\
$\gamma_{\rm B}$							&4.0		&4.2		&4.4		& 5.6		&7.5\\
$\gamma_{\rm exp}$\footnote{Taken from Ref.~\protect\onlinecite{MoritomoPRB}.}	
												&10		&10		&12		& 			&\\
$\gamma_{\rm exp}/\gamma_{\rm B}$		&2.5		&2.4		&2.7		& 			&\\
$W_{\rm exp}/W_{\rm B}$					&2.5		&			&2.2$-$2.8		&			&\\
\end{tabular}
\end{ruledtabular}
\end{table}

\section{Conclusion}

	We have investigated the electronic structure of Sr$_{2-x}$La$_x$FeMoO$_6$ by photoemission spectroscopy and LDA+$U$ band-structure calculations. A double-peak structure observed at about $-0.2$ eV (feature $A$) and $-1.3$ eV (feature $B$) was identified to be a Fe+Mo $t_{2g\downarrow}$ band and a Fe $e_{g\uparrow}$ band, respectively. The chemical potential shift due to electron doping was observed to be about 40$-$50 meV which was considerably smaller than the prediction of the band theory, 112 meV. Besides, the features $A$ and $B$ were both enhanced due to electron doping. We have pointed out that this unusual enhancement at the feature $B$ is probably indicating a characteristic distribution of doped electrons triggered by the strong Hund's rule energy stabilization in the 3$d^5$ configuration. From the observed band-narrowing, we have deduced a mass enhancement of $\sim$2.5 with respect to the band theory which is in good agreement with a specific heat measurement.

\begin{acknowledgments}
The authors would like to thank T. Kikuchi for technical support in the experiment. The experimental work has been done under the approval of the Photon Factory Program Advisory Committee (Proposal No. 00G011). This work was supported by a Grant-in-Aid for Scientific Research from the Japanese Ministry of Education, Culture, Sports, Science, and Technology.
\end{acknowledgments}


\begin{references}
\bibitem{note5}The accuracy of the location of A was $\pm10$ meV but that of B was much worse due to the broader feature. Thus we have also estimated the shift of the spectra using the steep edge between the features A and B, with accuracy of $\pm5$ meV.
\bibitem{note6}We note that the expression $[3d^5+4d^1]$ is a simplified description. A more realistic expression is [$3d^{5+\delta}+4d^{1-\delta}$] ($\delta \sim 0.5$) because M\"ossbauer measurements have concluded Fe$^{2.5+}$ (Ref.~\protect\onlinecite{Linden} and \protect\onlinecite{Nakamura}). However, this does not mean that the $[3d^5+4d^1]$ configuration does not exist at all; in terms of the local configuration-interaction description (around the Fe sites), $[3d^5+4d^1]$ still has a major weight and $[3d^6\underline{L}+4d^1]$, $[3d^6+4d^0]$, and $[3d^5\underline{L}+4d^2]$, etc. have some weight, respectively. (This is one of the reasons why the nominal valences of Fe and Mo are often written as 3+ and 5+ in spite of the M\"ossbauer measurements .) Thus we have picked up the major configuration $[3d^5+4d^1]$ for simplicity.
\bibitem{note1}If we assume that doped electrons may occupy the Fe 3$d$ states, Fe $t_{2g\downarrow}$ states, namely the feature A, should be enhanced due to electron doping and no enhancement would be observed at the feature B. This consideration further reinforces our argument.
\bibitem{note3}We note that the 80 eV curve is not shown because it is indistinguishable from the 90 eV curve.
\end{references}

\newpage 



\end{document}